# Cybermatter

*Daniel Stern*

*France Telecom – Research & Development*

*38 rue du General Leclerc, 92794 Issy-les-Moulineaux*

*daniel.stern@orange-ftgroup.com*

**ABSTRACT**

In this paper we examine several aspects of the impact of Cyberworld onto our Reality conceptions, and their social implications.

First we analyse the Communication as a principal component of Life: the information exchange cannot be dissociated from Life.

Second, we consider the progress in electronics which has enabled to transform any perceptible signal into a list of numbers to further store and transmit it. Trustable replication and transmission and almost-eternal conservation have given information contents and interactions a reliability and a consistency that belonged, beforehand, only to the tangible matter. This digital revolution laid the foundations of a new world: the world of information. Several characteristics fundamental to our Reality understanding are not longer applicable in this World: this one is made of a new thing.

In a third step we examine several aspects of this new thing and the impact of the Cyberworld onto our Reality conceptions. The Cyberworld matter, composed with information, could be called Cybermatter, characterised by Ubiquity and Dematerialization.

In a fourth and final step we discuss the use of the Cyberworld. It is a progress but beware of viewing it as only beneficial. All progress has a good and also an evil side. What will be the responsibility of the avatar's owners? Is there a risk to enlarge the "digital divide": on the one hand a mankind connected or connectable with/through a Cybeworld more and more attractive and populated; on the other hand another mankind out of the game.

**Keywords:** Life is Communication, Virtual vs. Real, Ubiquity, Dematerialization, New Matter

## FROM A NETWORK TO A NEW WORLD

Cyberworld. The Internet virtual world is no longer an SF fantasy. Its most important subset, the world wide web with billions of online documents, permanently updated, restlessly explored by hundreds millions of humans, is the most complex system ever created by mankind: its number of active connexions will soon approach those of the human brain ($10^{15}$).

J.-L. Borges had considered in his Aleph (Borges, 1956) that the book structure (juxtaposed leaves and volumes) had contributed to shape the human thought[1]. What if the information repository takes the form of a planet-sized web where all references are (nearly) immediately at hand and knowledge acquisition is a kind of navigation? And if we apply now this metaphor not only to searching static information but also to the inter-human relationships themselves, we guess the impact of this new communication structure on our reality representation and on our social behaviours.

---

[1] At least for those who can read.

The Cyberworld is not completed yet (will it be one day?). Nevertheless it is regularly used by many people for commercial purposes; game playing; social encounters; personal development; diary, literature and scientific publishing; file sharing, etc. For these people it is already the normal place to communicate with others.

A tool (telephone then data network) aimed at putting in relation remote people has taken such a substance that it is now called a World or a Space. In this paper we limit our scope to identifying a few crucial steps and also sketch what the Cyberworld is, its origins, its impact, and a few ideas on how it may change our conceptions of the real world.

## COMMUNICATION: A VITAL NEED

Communication is, without doubt, a principal component of Life. From the link maintaining together the cells of a simple Trichoplax colony to an Internet forum dedicated to Peace promotion, communication is one of the main competencies (s)elected by the billion[2] years of natural selection.

Natural selection favours the coordination between cells to let them associate/specialize in complex, multi cellular organisms because they have a better[3] gene survivability/transmission (Dawkins, 1976).

Closer to us in the life tree, communication is inherent to living beings having a sexual reproduction: it is obviously required to find (with different sophistication levels) a partner. For many species communication may also improve, thanks to a social cooperation, food searching in general and more precisely hunting and obviously agriculture. What is more, love parade, seduction behaviours, tend to form stable couples and then provide a better efficiency in gene transmission (Dawkins, 1976).

Social species are dominant with respect to isolated-individual species: the former tend to win space over the latter (same discussion as in note 3). And the social link is nothing but an inter-relation between individuals. A simple yet fascinating example is the coded dance a bee performs to warn its hive sisters about an interesting nectar site.

The societies are characterised by a permanent exchange of information between their individuals. This information exchange fulfils the only Life aim: multiply.

Within the Matter also exists a permanent exchange. But here it consists of energy, theoretically information-less[4]. So there is an important difference with Life. This latter is self-organised and an exchange between two living entities has a purpose: the emitter intends (consciously or, more often, without any conscious representation) to signal something to some*body*, and may even waits for a feedback.

Living beings, even the simplest living entities as genes, are self-organising because they multiply: Life is in growth, at the contrary of the Matter which is stationary (Corning 1998), (Barabasi 2002) as the proverb states about the matter "Nothing disappears, nothing appears, all transforms".

This information exchange cannot be dissociated from Life, the same manner the energy exchange is not dissociated from Matter[5]. This is an aspect of the classical mind/matter opposition, important to dualist philosophers, manicheists and Gnosis adepts. Closer to us, it is recalled by Bateson (1972) who opposes *Creatura* (living) to *Pleroma* (non living), after C.G. Jung, who himself borrowed it from Gnostic philosophers like Basilides of Alexandria.

Energy and Information are obviously related as the latter requires, to be elaborated, stored, transmitted, received, and finally interpreted, a certain amount of energy. And the job of researchers and engineers in many communication-related fields, for decades, consists of minimizing this amount.

---

[2] The oldest micro-organisms, found in Australia, are aged of more than 3 billions years.

[3] This point may be discussed as the virus and bacteria population is also enormous.

[4] There is still a debate in the Quantum Physics about the observer's role on the wave function collapse and therefore on the Reality (see Schrödinger's cat paradox); see also recent cosmological discussion on the "Anthropic" principle.

[5] Beside the matter-energy equivalence stated by Einstein's $E=mc^2$, a particle is defined in the Physics Standard Model by the *interaction* it represents.

Human symbolic capacity – and Life itself, supposed mankind is its highest manifestation as thought Teilhard de Chardin (1955) – is finally aimed at the same objective: how to represent and exchange more and more information without spending more energy (actually, by designing and utilizing sophisticated systems able to take the process in charge).

Actually if Shannon's information theory (Shannon, 1948) uses the entropy, a notion coming from thermodynamics to measure the information, it is indeed because the entropy is energy without work:

$$H(x) = -\sum_{i=1}^{n} p(i) \log p(i)$$

(Where H(x), entropy of system x, is defined as the sum, over all possible states i of x, of the probability p(i), multiplied by its inverse logarithm).

Concerning the human beings, communication operates as soon as a new-born[6] claims food. Actually this applies to all animal species where genitors positively respond to hunger signals received from their young (mammals, birds, many insects).

This primal relationship is at the core of what a young feels for its feeding parent, as well as the reverse relation where the feeding parent desires to please its young (Winnicott, 1958), (Klein, 1933). Hence it is the motivation and the background in which we learn our languages (body, gestural, spoken). This founds the expression mode for the desire.

The gap in communication skills that separates Human from other species is not only quantitative. The symbolic dimension has got, for humans, an exceptional development. Indeed, human beings are provided with, correlatively to spoken language, a huge symbolic representation ability, which, in addition and this is not its least characteristic, operates through several "levels". One (at least) of them is unconscious.

In this context thrives our almighty communication need, constantly pushing us to create more and more communication means and to (over)load them with symbolic content. This latter aspect is especially present in virtual reality.

Communicating is fundamental to Life. Mankind never will avoid this essential compulsion and many of the refinements we develop are aimed at this single objective: being able to communicate better, more often, and with less effort.

## EN ROUTE TO CYBERWORLD

In the second half of twentieth century the progress in electronics has given us the ability to transform any perceptible signal into a list of numbers, representing this signal more or less accurately according to the coding precision, to further store and transmit it. And with the error correction methods, all the technological[7] factors were set for the digital revolution. Trustable replication and transmission and almost-eternal conservation have given information contents and interactions a reliability and a consistency that belonged, beforehand, only to the tangible matter.

This digital revolution laid the foundations of a new world: the world of information.

In this world, several rules, yet obvious in the real world, soon, will be no more valid.

Whereas the telephone is somehow only an elaborate form of oral transmission, modern telecommunication networks are radically new. They bring means to materialize an immaterial content, and simultaneously to dematerialize into a digital content potentially any tangible object.

---

[6] What happens before birth between a mother and her child is also, without any doubt, a kind of communication.

[7] Technological factors as of course many other cultural and societal factors were already set. In his *1984* published in 1948, Georges Orwell imagines a perfect information system materialized merely on paper bills.

This gives birth to a new thing: a knowledge created, maturating, and circulating differently. A message may turn into a text or a speech and conversely. Opposing static to dynamic information becomes obsolete.

For example, a role playing game, a movie, a photo album, music (here, things are the most mature), texts, diaries are accessible, modifiable, through the very same interfaces with which we – sometimes even simultaneously – are talking with our fellows.

Soon we will be unable to distinguish "content", i.e. something prepared, written, stored in order to be later read, shown or listened to, from a dynamic message. After an era where the text had a social image, a cognitive impact clearly different from the speech we enter a new age where written and oral transmission modes may merge back.

In the technical jargon this is called the opposition synchrony/asynchrony. The mobile telephone, once designed to be the ultimate synchronous communication terminal (to allow us to speak everywhere and at any time) has become, thanks to the SMS (Short Message Service), the support of an asynchronous way to communicate and also the motivation for a new syntax. Similarly we can mention the Role Play Games which, in addition to the tightly synchronous interaction which is the initial motivation, are the reason to exchange numerous asynchronous messages (tactics, cheating…).

Other opposition currently disappearing is on the axis human/artefact. Software automata called bots often play the role of moderators to enforce the netiquette in the instant messaging chat rooms. How to differentiate a real living being from an automaton is an issue studied as soon as 1950 by Alan Turing in his famous test (Turing, 1950). In this test the subject, connected both to a real person and to a machine through terminals, has to determine, only by asking questions, which partner is human and which is the machine. Until now no machine has succeeded to fool the subject[8].

As soon as ambiguity is made possible, i.e., if any of us is replaceable with a clone, whatever its aspect, realistic or fantasy-like, an important and irreversible step will be passed in the inter-human communication. Undoubtedly, new social, political and philosophical issues will be posed: rights, duties, liberties of these virtual creatures.

As for many human activities (consistently with what we said above), business and relations are the most stringent motivation to Internet development. Economics are even the religious basis to certain Internet prophets: Negroponte (1995) forecasts that soon the *matter* trade will yield to *information* trade and that "bit streams" volume will pass "atom streams" volume. His prediction may realize as we will see below.

# LAND!

The word "cyberspace" was popularised in the 1980s by William Gibson, the Canadian science fiction writer, in his novel Neuromancer. He defines it as a "Consensual hallucination experienced daily by billions of legitimate operators (…)". More technologically we can define it as:

- firstly, a set of technologies aiming at representing and reproducing real, virtual or mixed elements;
- secondly, a set of technologies to let humans access these elements;
- thirdly, and by metonymy, the perceived result of all the corresponding processes, i.e., the real, virtual or mixed elements themselves: a sort of world.

The Cyberworld shares several characteristics with the Real world. Like the real world its exhaustive cartography is impossible; its existence is subject to speculation. Nevertheless we may precise several crucial characteristics. As we saw in previous section several old distinctions have started to disappear. At the limit, new effects are expected. What are the main features of this new (found) land?

---

[8] A common game in the chat rooms consist of finding which persons are bots by asking them complicated questions

## Virtuality

Most of the references to the Cyberspace highlight its virtual reality aspect, and sometimes reduce it to this one.

Mixed-reality systems bring interesting precisions. A good illustration of the distinction and the superposition real/virtual is the example of an archaeologist exploring excavations, who gets in his goggles additional information in superimposition to the examined items.

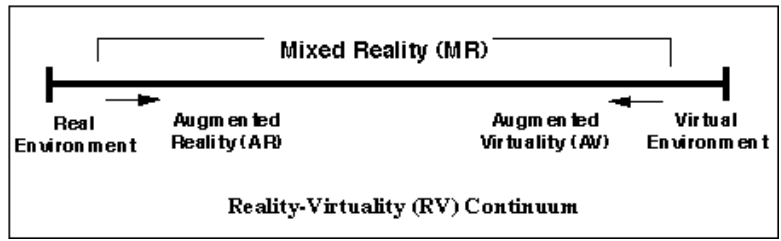

*Figure 1: Milgram's Reality-Virtuality Continuum*

The technology creates a continuum between real and virtual by introducing intermediate steps (see figure extracted from (Milgram, 1994).

The notion of virtuality is difficult and when defined, it is merely opposed to reality – as in Milgram's Continuum –, and often associated to imagination (references). Numerous conferences We do not attempt to redefine it. Only let us notice that it relies – technically and intellectually – on at least two main factors: dematerialization and ubiquity of the contents.

### *A new Thing*

Digitalizing content (images, sounds, texts) makes it more available and also allows more people simultaneous access to it. Besides, anyone may personalize and adapt this content and/or its presentation to his/her preferences.

A huge knowledge becomes can be reached by anyone. This is a new kind of knowledge, dynamical and volatile, subject to endless annotating and modifying. This breathtaking Babel Library, of which Borges saw the immensity but not the dynamicity (Borges, 1998) is a living organism, permanently created by us and subject to Darwin natural selection laws.

Many complain that the new knowledge is incomplete, often false. We will not discuss here this issue, except by remarking that the approval process through which new ideas were accepted should be replaced with networked, "moderated" approval.

Since personal content may be put on line, then personal data, the "profile" is accessible and modifiable by other people. Many people have several pseudonyms, each one with possibly different age/sex/city… Many contents are already built dynamically, by "syndication" using links to news or other media. Why could not my avatar be subject to a same volatility, dependent on my mood, or the day-to-day fashion?

Probably, besides a stable administrative identity, submitted to a strict authentication, we will have more identities, changing and not guaranteed. This is consistent with the extreme ideology which criticizes the notion of citizenship (EFF, 1996). In order to protect one's identities from wrongdoers there is a strong demand for security. Identity theft has not waited for Cyberspace to exist. Simply, there will be new opportunities and also new protections.

### *Ubiquity*

Abolishing the distances leads to ubiquity. This is possible due to two reasons. On the one hand human interfaces (audio, video, text, gesture, haptic, etc.) are distributed and always on – actually capable to autonomously manage their availability – where I am right now, constituting what is now called a *pervasive* environment. On the other hand end-to-end, secure, reliable and rapid connectivity is provided by the network.

The impression of presence depends, of course, on the quality of the signal, which itself relies on the input and output interfaces and on the transmission channel.

In this situation called ubiquitous or pervasive communication two individuals are always able to communicate whatever distance separates them.

Ubiquity is already on, brought by modern telecommunications which were initially motivated by this need: "I hear you as if you were here!". It involves de-location: the first common question when speaking through a mobile phone is "where are you?".

*A fortiori* when even the pieces of interface are anywhere, get localized and localize the part of environment currently transmitting/supporting the communication will be a challenge.

Finally, ubiquity defines a specific topology as we indicate below.

**Time, Memory and Observer**

The virtuality of the Cyberworld is a consequence of the two previous factors (new space, new thing). But to give it substance, it still needs a third important term: an observer's cognition. And thought, because it is a time-related story, gives coherence to the rest.

Here intervenes an important aspect, which is not specific to virtual worlds: the hypertext navigation. The hypertext navigation composes a narration with a chronological thread possibly different for two observers. It becomes almost impossible that both share a common reference frame. Beforehand, this was called "two readings of a same story". Now, the stories are themselves objectively different!

Our thought process is itself "hyper textual": it is constituted with several parallel threads linked at some steps by (recalled or actual) images, sounds, impressions, etc. The hypertext navigation can be analysed as a successful attempt to externalise our internal process, but based on different operations. This property already studied (Rouet, 1996) will not be discussed deeper here. But it is important for our purpose: the Cyberworld may function like the Human mind and reversely the Human mind may find in the Cyberworld an obvious, almost natural, extension.

## CYBERMATTER

Defining Reality is a highly complex problem that has been dealt with by philosophy and science for thousands years, probably since the Human has been able to think properly. The purpose here is not to summarize what the main conceptions are, from Heraclites and Buddha to the Quantum Physics, via different forms of Idealism. We would like to examine the impact of Cyberworld potential existence onto our Reality conceptions.

Most of these conceptions grant that the World is Space containing some Matter. Cyberworld is also a kind of Space containing a kind of *Matter*. This latter, composed with information could be called *Cybermatter*.

Ubiquity and Dematerialization are two faces of a fundamental characteristic of this Cybermatter. Contrary to Real Matter objects, a Cyberobject is not necessarily unique neither – which is a consequence – uniquely located. As it is dematerialized it may be anywhere; reciprocally to be anywhere, an object, necessarily[9], cannot be material. An environment which seems real in the Cyberworld may be completely forged. Indeed, Cyberworld resembles a fantasy kingdom or the Unconscious.

Most of human important breakthroughs have had their feedback on our World understanding: the Relativity, Quantum Physics, Psychoanalysis. It is then probable that the Cyberworld impact several concepts under discussion for many centuries about the Real World.

Among these concepts let us cite the uniqueness. A real object is unique because I, as an observer, think it unique. A Cyber object is ubiquitous because I, as an observer, conceive it as ubiquitous.

---

[9] This question may be discussed in the frame of Quantum Physics considerations on non-locality

What is more, as an object among others, an observer is also dematerialized and may be ubiquitous: *other observers may be avatars of other people or even of myself!* Even if my identity is subject to a strong and inviolable authentication, I will be able to, from time to time, send in the Cyberworld, several agents acting as me.

The observer's uniqueness is then concerned and therefore also his/her memory, personal narration, along the time, as mentioned above. Reversely, common views on uniqueness, thought, cognition in the real world may be impacted (these points will not be discussed here).

We may summarize these concepts by the following scheme.

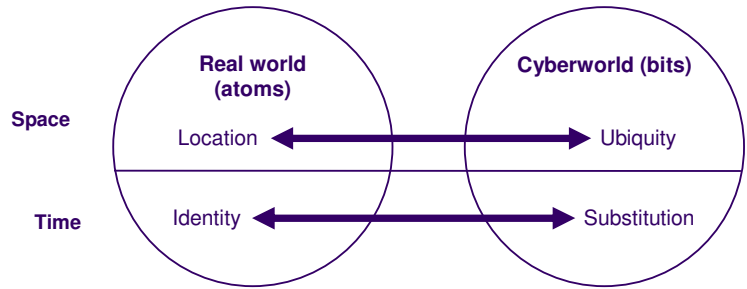

*Figure 2: Cyberworld vs. Real World*

In this figure, fundamental characteristics of the Real world are represented with their equivalent in the Cyberworld. Time, in both cases, is not pre-existing: it follows from the existence of a Space-based memory. Several authors have begun to deal with these topological considerations (Deutsch, 1998).

## WHAT IS THE USE OF CYBERWORLD?

What are the advantages of abolishing the distances, immersing in virtual spaces, dissolving one's identity, and replicating one's appearance? As we tried to explain above, these questions address very fundamental aspects. Communication is an attribute of Life. Then telecommunications, with their most recent development, i.e., the Cyberworld, are parts of the human evolution. Therefore the Cyberworld is a progress, like the air transports, the video broadcast, the medical breakthroughs, etc. But, contrary to these latter, it concerns the Communication and therefore may have important societal and psychological impacts.

Cyberworld is a progress but beware of viewing it as only beneficial. All progress has a good and also an evil side. Even if the drawback is not cataclysmic many people may be left behind. As for this, we disagree with a certain angelic optimism as the one of the Electronic Frontier Foundation who sees in the Cyberspace a Promised Land for the persecuted, a place where it is good to live, at least better than in our Real World: « *Governments of the Industrial World, you weary giants of flesh and steel, I come from Cyberspace, the new home of Mind. On behalf of the future, I ask you of the past to leave us alone. You are not welcome among us. You have no sovereignty where we gather* » (EFF, 1996).

When ubiquity and no-location are normal, and when identity and property are replaced or mixed with multiplicity, to what extent avatars will act as our agents? How to limit deviant behaviours? What will be the responsibility of the avatar's owners?

Several authors (Suler, 1996, 1998), (Greenfield, 2006) have addressed the Cyberspace impacts on social, moral and ethics. Generally they conclude that, although there exist several autistic or deviant behaviours, there is no major social difference between virtual and real worlds. Furthermore this may lead that there is no distinction real/virtual.

Let us grant that a communication everywhere and anytime will enable to share more, to have access to a broader knowledge, to visit far-away or disappeared places and times, and also to profit from business opportunities. If we add new technologies discovered or improved on the way: new interfaces, aware context, speech, face or gesture recognition and synthesis, immersive and mixed-reality systems, then we may expect that the final account will be highly positive.

At the opposite, as often feared, there is a trend to isolation: indeed, why would the virtual urge us to encounter one's fellows in the real world? Correlatively, there is a possibility that the investment in the Real World became economically less worthy than in the Cyberworld (Negroponte, 1995). This would enlarge the "digital divide": on the one hand a mankind connected or connectable with/through a Cybeworld more and more attractive and populated; on the other hand another mankind out of the game. The derealization foretold[10] by Baudrillard (1981) would happen, with a bad effect for the majority of the mankind.

In the question "is the Cyberworld a *real* progress and if yes, to whom, and for what?" the term "real" is somewhat paradoxical and this is not only a play on words: maybe the Cyberworld is not directly a progress for the real world, because it is a new world.

Indeed, the Human species, stuck on a planet now entirely explored, have begun to build *ex nihilo* an artificial and modifiable Terra Incognita. Everyone will be able to project onto this New World his/her fantasies, hopes, phobias, passions. Already land of opportunities it is bound to become a stake of power and then of rivalries. And, as in the Real world, protection laws are mandatory to enforce fair behaviours.

Abolishing the distance especially addresses a very important and archaic issue as mentions Michel Serres (2001) who sees in this "delocalization of the neighbourhoods and of the cognitive values" a major step in the human evolutionary process, comparing it to the discovery of the fire.

If it allows humanity to discover, meet (at first in virtual then in real life) other people, their culture and emotions, surroundings, is it not a worthwhile progress?

---

[10] See also *eXistenZ*, a Cronenberg's movie, where a game in a multiverse Cyberworld ends up by a fierce battle between Reality supporters and opponents.

# References

Barabasi, L. (2002), Linked: The New Science of Networks, Perseus, Cambridge.

Bateson G. (1972) Steps to an Ecology of Mind, The University of Chicago Press, 1972.

Baudrillard J. (1981) Simulacres et Simulation, Galilée, Paris.

Borges, J.-L. (1956) Ficciones. Sur, Buenos Aires.

Dawkins, R. (1976, 1989, 2006) The Selfish Gene, Oxford University Press, Oxford.

Deutsch, D. (1998) The Fabric of Reality, Penguin Books Ltd, London.

EFF (Electronic Frontier Foundation) (1996) http://homes.eff.org/~barlow/Declaration-Final.html

Greenfield, A. (2006) Everyware: The Dawning Age of Ubiquitous Computing. New Riders Publishing.

Klein M. (1933) The early development of conscience in the child. In Love, guilt and reparation and other works 1921-1945. The writings of Melanie Klein (Vol. 1, pp. 248-267), Hogarth, London.

Milgram P. et al. (1994) Augmented Reality: A Class of Displays on the Reality-Virtuality Continuum, SPIE Vol. 2351, Telemanipulator and Telepresence Technologies.

Negroponte N. (1995) Being Digital. Alfred A. Knopf, Inc., New York.

Rouet J.-F. & al (1996) Hypertext and Cognition, Lawrence Erlbaum Associates.

Serres M. (2001) Hominescence. Editions Le Pommier, Paris.

Shannon C. (1948), A Mathematical Theory of Communication, the Bell System Technical Journal, Vol. 27, pp. 379–423, 623–656, July, October.

Suler, J.R. and Phillips, W. (1998) The Bad Boys of Cyberspace: Deviant Behavior in Multimedia Chat Communities. CyberPsychology and Behavior.

Suler J.R. Cyberspace as Psychological Space, http://www.rider.edu/~suler/psycyber/psycyber.html

Teilhard de Chardin P. (1955) Le Phénomène Humain, Editions du Seuil, Paris.

Turing A. M. (1950) Computing Machinery and Intelligence, Mind, Vol. 49, pp. 433-460.

Winnicott, D.W. (1958). Collected Papers. Through Paediatrics to Psycho-Analysis., London: Tavistock Publications.